\documentclass[12pt,oneside,english,american]{amsart}
\usepackage[T1]{fontenc}
\usepackage[utf8]{inputenc}
\usepackage{url}
\usepackage{amstext}
\usepackage{amsthm}
\usepackage{amssymb}
\usepackage{geometry}
\usepackage{setspace}
\makeatletter

\let\SF@@footnote\footnote
\def\footnote{\ifx\protect\@typeset@protect
    \expandafter\SF@@footnote
  \else
    \expandafter\SF@gobble@opt
  \fi
}
\expandafter\def\csname SF@gobble@opt \endcsname{\@ifnextchar[
  \SF@gobble@twobracket
  \@gobble
}
\edef\SF@gobble@opt{\noexpand\protect
  \expandafter\noexpand\csname SF@gobble@opt \endcsname}
\def\SF@gobble@twobracket[#1]#2{}

\numberwithin{equation}{section}
\numberwithin{figure}{section}

\usepackage{babel}

\makeatother

\usepackage{babel}
\begin{document}
\title{\selectlanguage{english}%
Short Title: Gadgil's Theory and Models in Ecology}

\maketitle

\section*{\textbf{Full Title:} From life-history model to theories of Eeological diversity: The impact of Madhav Gadgil's work\foreignlanguage{english}{\protect}\protect\footnote{Guest Editors' Perspectives cum Mini-review article for \textit{Journal
of Theoretical Biology}, VSI: Ecological Modelling: in honour of Prof. Madhav Gadgil}}

\vspace{0.7cm}

\begin{center}
\author{\selectlanguage{english}%
{\Large Arni S.R. Srinivasa Rao},}
\begin{center}
Laboratory for Theory and Mathematical Modeling, 
\par\end{center}

Division of Infectious Diseases - Department of Medicine, 
\begin{center}
Medical College of Georgia, 
\par\end{center}

\begin{center}
Department of Mathematics, 
\par\end{center}

\begin{center}
Augusta University, Georgia, U.S.A. 
\par\end{center}

\begin{center}
Email: arrao@augusta.edu 
\par\end{center}

\begin{center}
\vspace{0.4cm}
 
\par\end{center}
\author{\selectlanguage{english}%
{\Large Harman Jaggi}, }

Ecology and Evolutionary Biology,

Princeton University, Princeton, U.S.A.

Email: harmanj@princeton.edu 
\begin{center}
\vspace{0.4cm}
 
\par\end{center}
\author{\selectlanguage{english}%
{\Large Shripad Tuljapurkar},}

Department of Biology,

Stanford University, Stanford, U.S.A

Email: tulja@stanford.edu

\end{center}
\begin{abstract}
In this article, we review and provide perspectives on the foundational contributions of the renowned ecologist Madhav Gadgil, including the Gadgil-Bossert life-history model, ecological diversity, population biology, harvesting theory, human-ecological systems, co-evolution, adaptive management, and environmental democracy. We take the Gadgil-Bossert model of reproductive allocation as a starting point for reviewing Gadgil's broader contributions to evolutionary and population theory. Then we examine how density regulation shapes selection on reproductive allocation in his work on $r$- and $K$-selection. We then consider how interactions among genotypes determine whether a mixture can remain stable and resist invasion. We outline the key theoretical developments arising from these works and discuss their continued influence on contemporary ecological thought and present-day ecological systems. 
\end{abstract}


\begin{center}
\vspace{0.2cm}
 
\par\end{center}

\section{\textbf{Introduction}}

Life history for a species is viewed by ecologists in terms of the
distribution of survival and reproduction over an organism's life
cycle \cite{stearns1976life,jaggi2024density}. The fundamental ideas
started with Lotka who developed the mathematical formulation of demography in the early twentieth century (summarized in Lotka's 1939 book, reprinted in 1998), showing the importance of the stable age structure and the long-run growth rate $r$. \cite{Fish} complemented this picture by discussing the importance of reproductive value, a particularly important metric that shows how and organism's contribution to future population changes with age. These concepts began a theoretical and empirical journey that attempts to span any kind of organism with any kind of life cycle.

A fundamental idea in life history theory is that $r$ serves as fitness in life history theory (an idea only validated by population genetics models in the 1970s). Using fitness as measured by $r$, there was ``intense theoretical activity'' between 1950 and 1973 (as nicely stated by Stearns in his 1976 review \cite{stearns1976life}). Much of this work also realized that $r$ needed to be optimized under biological and energetic constraints, not merely maximized. The famous paper by Gadgil and Bossert \cite{GA-Bos1970} laid out this concept and traced many of its implications.

From a biological point of view, of course, a central issue is to
enumerate predictions made by the theory, so that these predictions
can be tested. The Gadgil-Bossert analysis examined optimum allocation between reproductive output, maintenance and survival, an idea that underlies much life history theory. In their best-known analyses, they suppose that optimal allocation depends on an individual's residual reproductive value and so varies with age. They predict that when residual reproductive value is high, early in life, survival and future reproduction are favored relative to current reproduction. The allocation reverses at late ages, as residual reproductive value drops, so the analysis predicts that allocation to reproduction is more important than allocation to survival.

The Gadgil-Bossert approach remains influential in the development
and application of life history theory, especially to organisms with
complex long life cycles. For a notable example that applies that
approach to the long lives of humans, Kaplan et al. present models
and data to show that a long human life span co-evolved with a longer juvenile period, age-dependent increase information processing and storage, and intergenerational resource flows \cite{kaplan2002emergence}. This human life cycle work has been important in the field of aging.

The subsequent sections of this mini-review summarize some of these
mathematical developments. We take the Gadgil-Bossert model of reproductive allocation \cite{GA-Bos1970} as a starting point for reviewing Gadgil's broader contributions to evolutionary and population theory. We first examine how density regulation shapes selection on reproductive allocation in his work with Solbrig on $r$- and $K$-selection \cite{Ga-Solb}. We then consider how interactions among genotypes determine whether a mixture can remain stable and resist invasion \cite{GA-VN_GA1980}. The final sections turn to population persistence and predictability. We review how refugia can help exploited populations persist under harvesting \cite{NVJ-MG1990} and how dispersal and spatial scale
affect our ability to forecast population dynamics \cite{Xuetal2005,gadgil1971dispersal}.

\section{\textbf{The Gadgil-Bossert Life-History Model}}

The life-history model \cite{GA-Bos1970} is about understanding how
organisms maximize their fitness, where fitness is measured in terms
of lifetime reproductive success. The life-history is understood by
dividing age into discrete classes $0,1,2,...\omega$ \cite{tuljapurkar2025generalizing}.
Here $\omega$ is the maximum age. Fitness is typically measured by
taking into account age at first reproduction, growth, mortality risk,
and future reproductive capacity. Three central assumptions that govern
the Gadgil-Bossert model are: $i)$ an organism has limited energy
to be allocated among reproduction, growth, and survival, $ii)$ increased
allocation of energy to reproduction reduces the chance of future
survival and growth of organisms, $iii)$ the chance of dying increases
due to increase in reproductive effort. They provide an explicit optimization
framework in the life-history to understand the profit of present
reproduction versus the cost to future survival and capacity to produce
offspring (fecundity).

Gadgil-Bossert first derived an equation of the form 
\begin{equation}
\sum^{\omega}_{x=0}e^{-mx}l_{x}b_{x}=1,\label{eq:1}
\end{equation}
where $m$ is the Malthusian parameter, $l_{x}$ is the probability
of survival from birth to age $x$, $b_{x}$ is the number of offspring
produced at age $x$. The quantity $m$ measures fitness. Without
considering the population density, Fisher, in his theory of natural
selection \cite{Fish}, showed that $m$ will increase by assuming
fixed degree of inbreeding in a constant environment. The survival
function $l_{x}$ was expressed by 
\begin{equation}
l_{x}=\prod^{x-1}_{i=0}\theta_{i}f_{1}(\alpha_{i})g_{1}(\beta_{i})h_{i},\label{eq:2}
\end{equation}
where $\theta_{i}$ is the maximal survival probability from age $i$
to $i+1$ in a non-limiting, predator-free environment, $\alpha_{i}$
is the reproductive effort of an individual at age $i$, $\beta_{i}$
is the fraction of the utilizable resources available to an individual
at age $i$, $h_{i}$ is the probability of escaping death through
predation at age $i.$ We note $f_{1}(\alpha_{i}),g_{1}(\beta_{i})\in[0,1].$
It was assumed that $f_{1}$ is a monotonically decreasing function
i.e. reproductive effort reduces survival (i.e. $f_{1}'(\alpha_{i})<0),$
and $g_{1}$ is a monotonically increasing function i.e. resource
availability improves survival (i.e. $g_{1}'(\beta_{i})>0).$

Let $s_{x}$ be a function representing the population size at age
$x.$ The function $s_{x}$ is expressed by 
\begin{equation}
s_{x}=s_{0}+\sum^{x-1}_{i=0}\delta_{i}f_{2}(\alpha_{i})g_{2}(\beta_{i}),\label{eq:3}
\end{equation}
where $\delta_{i}$ is the increment in size of individuals from age
$i$ to $i+1$ who are making no effort in reproduction. The function
$f_{2}$ is monotonically decreasing, and the function $g_{2}$ is
monotonically increasing, both in the interval $[0,1].$ The quantity
$b_{x}$ in (\ref{eq:1}) is represented by 
\begin{equation}
b_{x}=s_{x}f_{3}(\alpha_{x})g_{3}(\beta_{x}),\label{eq:4}
\end{equation}
where $f_{3},g_{3}\in[0,1]$, and $f_{3}'(\alpha_{x})<0$, $g_{3}'(\beta_{x})>0$.

The life-history model in {[}\ref{eq:1}{]} resembles characteristic
equation of the Euler-Lotka population integral equation over a continuous
age $x$ represented in 
\begin{equation}
\int^{\infty}_{0}e^{-ra}l(a)m(a)da=1,\label{eq:4a}
\end{equation}
where $r$ is called the intrinsic rate of population growth, $l(a)$
is the survival probability of an individual reaching age $a,$ and
$m(a)$ is the maternity function at age $a.$ Taking the first-order
and the second-order derivatives of the integral in (\ref{eq:4a}),
we see that 
\begin{align}
\frac{d}{dr}\left(\int^{\infty}_{0}e^{-ra}l(a)m(a)da\right) & <0,\nonumber \\
\frac{d^{2}}{dr^{2}}\left(\int^{\infty}_{0}e^{-ra}l(a)m(a)da\right) & >0.\label{eq:4b}
\end{align}

\subsection{The Life-History Model Optimization Problem}

The optimization problem of life-history model in the absence of future
survival after $\omega$ is expressed as 
\begin{equation}
\max_{\alpha_{i}}\left(\alpha_{0},\alpha_{1},...,\alpha_{\omega}\right),\label{eq:5}
\end{equation}
subject to 
\[
0\leq\alpha_{i}\leq1,\alpha_{\omega}=1.
\]
The optimal $\alpha_{k}$ satisfies 
\begin{equation}
\frac{d}{d\alpha_{k}}\left[P_{k}(\alpha_{k})-C_{k}(\alpha_{k})\right]=0,\label{eq:6}
\end{equation}
where $P_{k}$ is defined as the profit function at age $k$, such
that 
\begin{equation}
P_{k}(\alpha_{k})=l_{k}b_{k}(\alpha_{k})\text{ and }\frac{dP_{k}}{d\alpha_{k}}>0.\label{eq:7}
\end{equation}
Here $C_{k}$ is cost function arising from decreased survival and
growth at later ages. The function $C_{k}$ is represented by 
\begin{equation}
C_{k}(\alpha_{k})=F_{k+1}(\alpha_{k}=0)-F_{k+1}(\alpha_{k}),\label{eq:8}
\end{equation}
where $F_{k+1}(0)$ is the total fitness contribution by the ages
greater than age $k$ for $\alpha_{k}=0,$ and $F_{k+1}(\alpha_{k})$
is the total fitness contribution by the ages greater than age $k$
for $\alpha_{k}>0.$ Gadgil and Bossert argued that the big-bang reproduction
occurs when $\alpha_{i}=0$ for all $i<k$ and $\alpha_{k}=1,$ and
using a Malthusian parameter $m_{k}$, they considered optimization
of $e^{m_{k}}$ using an appropriate choice of $k,$ where $e^{m_{k}}$
was defined by 
\begin{equation}
e^{m_{k}}=\left(l_{k}b_{k}\right)^{1/k},\label{eq:9}
\end{equation}
where 
\[
l_{k}=\prod^{k-1}_{i=0}\theta_{i}\beta_{i}h_{i},\text{ and }b_{k}=\left(s_{0}+\sum^{k-1}_{i=0}\delta_{i}\beta_{i}\right)\beta_{j}.
\]
The life-history model predicts that breeding occurs repeatedly if
$P_{k}$ is convex or $C_{k}$ is concave. Otherwise, breeding occurs
only once.

\section{\textbf{The $r-$ and $K-$ Selection Theory}}

The discussion on the relationship and mutual contribution due to
$r$ (the intrinsic growth rate) and $K$ (the carrying capacity)
is classical in population biology and demography. Gadgil, in his
article jointly with Solbrig in \cite{Ga-Solb}, showed that the value
of $r$ for a population does not depend solely on the birth rate
of the population. They suggest that the true selection of values
of $r$ and $K$ depends on the allocation of resources between reproduction
and non-reproductive functions.

Let $r(f(N))$ be the per-capita growth rate of a population of size
$N$ with density $f(N)$. Then $r(f(N))$ can be expressed as 
\begin{equation}
r(f(N))=b(f(N))-d(f(N)),\label{eq:3.1}
\end{equation}
where $r(0)$ can be considered as intrinsic rate of increase, $b(f(N))$
and $d(f(N))$ are birth rate and death rate as functions of density
$f(N).$ Next, we will show how the birth rate alone is not enough
or relevant for $r(0).$

\subsection{Density Dependent Versus Density Independent Mortality}

Let $K_{1}$ and $K_{2}$ be two equilibrium densities such that $K_{1}<K_{2}.$
An increase in density-independent mortality implies that the quantity
$d(f(N))$ increases, which causes the equilibrium density shifts
$K_{1}$ to $K_{2}$ from below. The quantities $b(f(N))$ and $d(f(N))$
increase at equilibrium due to the relaxation of density-dependent
regulation. Hence, higher $b(f(N))$ alone does not necessarily assist
in the selection of $r(0).$

\subsection{Resource Allocation Strategies}

Let $\varphi\in[0,1]$ be the fraction of resources required for reproduction.
There are two kinds of strategies in resource allocation and maximizing
the quantity $r(0).$ The strategies for $r(0)$ involve maximizing
$r(f(N))$ at low $f(N)$, higher $\varphi,$ lower allocation to
competitive and survival traits. The strategies for $K$ involve maximizing
$r(f(N))$ at high $f(N)$, higher allocation non-reproductive traits
enhancing survival. These two types of strategies are comparative
but not absolute. Gadgil and Solbrig argue that higher density-dependent
mortality indicates a higher optimal $\varphi.$ A straightforward
extension of (\ref{eq:3.1}) using $\varphi$ could be 
\begin{equation}
r(f(N),\varphi)=b(f(N),\varphi)-d(f(N),\varphi),\label{eq:3-2}
\end{equation}
where $\varphi\rightarrow1$ implies $r-$selection, and $\varphi\rightarrow0$
implies $K-$selection.

\subsection{Standard Logistic and Fisher-Kolmogorov-Petrovsky--Piskunov Logistic
Growth Models}

Under the classic $r-K$ selection framework, the population growth
model is represented by 
\begin{equation}
\frac{dN}{dt}=rN\left(1-\frac{N}{K}\right).\label{eq:3-3}
\end{equation}
The solution $N(t)$ for the model (\ref{eq:3-3}) is 
\begin{equation}
N(t)=\frac{K}{1+N(0)\exp(-rt)},\label{eq:3-4}
\end{equation}
where $N(0)$ is the initial population (for instance, see \cite{Raotextbook,Kapurtextbook}).
The two situations $N\rightarrow0$ and $N\rightarrow K$ can be interpreted
within the life history optimization framework through $\varphi.$
The higher $\varphi$ corresponds to $r-$selection, and the lower
$\varphi$ corresponds to $K-$selection. The situation $N\rightarrow0$,
corresponds to a situation of maximizing $r$, and the situation $N\rightarrow K$
corresponds to effectively maximizing $K$. The logistic growth model
also suggests that $r-K$ selection is shaped by population structure
and environmental constraints.

A PDE generalization of the standard logistic growth model using a
diffusion coefficient $D$ and the population density $u(a,t)$ with
initial density $u_{0}(a)$ is represented by 
\begin{equation}
\frac{\partial u}{\partial t}=D\frac{\partial^{2}u}{\partial a^{2}}+ru\left(1-\frac{u}{K}\right).\label{eq:3-5}
\end{equation}
The model in (\ref{eq:3-5}) is referred to as Fisher-Kolmogorov-Petrovsky-Piskunov
or Fisher-KPP equation \cite{CaiFisherKPP,GirardinFisher-KPP,Griff-FisherKPP,KolmororovFisher-KPP}.
The Fisher-KPP equation is also represented through 
\begin{equation}
\frac{\partial u}{\partial t}-D\frac{\partial^{2}u}{\partial a^{2}}=au-bu^{2},\label{eq:3-6}
\end{equation}
where $\frac{a}{b}$ is the carrying capacity. This type of model
(\ref{eq:3-6}) is also referred to as diffusive logistic model \cite{Du-F-KPP,AlvarezFisher-KPP}.
Such traveling wave equations have been widely used in understanding
properties of logistic growth, spatial movement through diffusion
coefficient, and invasion biology. The Fisher-KPP model is a natural
extension to study spatial dynamics properties in the standard $r-K$
ODE model studied previously. The wave speed $v$ can be seen to follow
the relation $v\geq2\sqrt{rD}.$ The equations of the type represented
in (\ref{eq:3-6}) are studied in a fixed habitat or infinite habitat
using the nonlocal diffusion equation such as 
\begin{equation}
\frac{\partial u}{\partial t}=\int_{\mathbb{R}^{n}}J(x,y)u(y,t)dy-u+ru\left(1-\frac{u}{K}\right),\label{eq:3-7}
\end{equation}
where $J:\mathbb{R}^{n}\times\mathbb{R}^{n}\rightarrow\mathbb{R}$,
and for spatially varying $r$ and $K$, the heterogeneous reaction-diffusion
systems are represented by 
\begin{equation}
\frac{\partial u}{\partial t}=\nabla.\left(D(a)\nabla u\right)+r(a)u\left(1-\frac{u}{K}\right).\label{eq:3-8}
\end{equation}

\section{\textbf{Population-Genetic Evolutionary Game Theory}}

Gadgil, along with S. Gadgil (his wife and collaborator) and V. Nanjundiah
in \cite{GA-VN_GA1980}, extended the classical evolutionarily stable
strategy proposed by Smith and Price \cite{Smith-Price1970}, and
developed a dynamic population-genetic theory of density-dependent
selection. Using the concept of a stable mixture of genotypes in evolutionary
dynamics, Gadgil and others introduced the notion of evolutionarily
stable composition. They obtained results on the existence of steady-states
in stability analysis, invasion resistance of mutant genotypes, and
provided a local perturbation analysis. When a population consists
of $n$ asexually reproducing genotypes, they show, the number of
steady states can be as large as $2^{n}-1.$ Two-genotype case and
three--genotype case are studied. The two-genotype case coincides
with Smith and Price's evolutionary stable strategy.

Let $p_{i}$ be the frequency of the $i^{th}-$genotype for $i=1,2,...,n$.
The mathematical model describing the dynamics of $p_{i}$ as considered
in \cite{GA-VN_GA1980} was

\begin{equation}
\frac{dp_{i}}{dt}=p_{i}(m_{i}-\bar{m}),\label{eq:4-1}
\end{equation}
where $p_{i}\geq0,$ $\sum_{i}p_{i}=1,$ $m_{i}$ is the $i^{th}$
genotype's fitness (Malthusian fitness), which is expressed as $m_{i}=\frac{1}{N_{i}}\frac{dN_{i}}{dt}.$
Here $N_{i}$ is the size of the $i^{th}-$genotype. The quantity
$\bar{m}=\sum_{i}m_{i}p_{i}$ is the mean population fitness. At steady-state,
$p_{i}(m_{i}-\bar{m})=0$ for all $i.$ Hence, $p_{i}=0$ or $m_{i}=\bar{m}$
for each $i.$ This implies that, at steady-state, a subset of genotypes
(say, $k$ subtypes) coexist with equal fitness.

Let $A_{ij}$ be the payoff to the $i^{th}$ genotype when interacting
with the $j^{th}$ genotype, measured in fitness units. When the population-genetic
dynamics were considered, the Malthusian fitness was expressed as
\begin{equation}
m_{i}=M(0)+\mu_{i},\label{eq:4-2}
\end{equation}
where $M(0)$ is the baseline fitness and $\mu_{i}$ is the mean payoff
from social interactions, and is represented by 
\begin{equation}
\mu_{i}=\sum^{n}_{j=1}A_{ij}p_{j}.\label{eq:4-3}
\end{equation}
The model in (\ref{eq:4-1}), using $A_{ij}$ can be now expressed
as 
\begin{equation}
\frac{dp_{i}}{dt}=p_{i}\left(\sum_{j}A_{ij}p_{j}-\sum_{j,k}A_{jk}p_{j}p_{k}\right).\label{eq:4-4}
\end{equation}
The model (\ref{eq:4-4}) generates trajectories in genotype-frequency
space. A population satisfies the evolutionarily stable composition
mentioned above, if it satisfies stability against small internal
perturbations of genotype frequencies and invasion resistance against
any immigrant genotype. A quantity $\bar{p_{i}}$ is considered in
the perturbation analysis on $p_{i}$ by defining $\bar{p_{i}}=\frac{D_{i}}{D}$,
where 
\begin{equation}
D=\left|\begin{array}{cccc}
1 & 1 & \cdots & 1\\
A_{11}-A_{21} & A_{12}-A_{22} & \cdots & A_{1k}-A_{2k}\\
\cdots & \cdots & \cdots & \cdots\\
A_{11}-A_{k1} & A_{12}-A_{k2} & \cdots & A_{1k}-A_{kk}
\end{array}\right|_{k\times k},\label{eq:4-5}
\end{equation}
and $D_{i}$ is obtained from $D$ by replacing the first elements
of the $i^{th}$ column of the determinant in (\ref{eq:4-5}) by $1$
in the $i^{th}$ row and $0$s in the other rows such that $D=\sum^{k}_{i=1}D_{i}.$
If a small perturbation $\delta p_{i}$ is imposed on $\bar{p_{i}}$,
i.e. $p_{i}=\bar{p_{i}}+\delta p_{i}$ for $i=1,2,...,k$ and $\delta p_{i}\ll\bar{p_{i}}.$
Substituting these perturbations in (\ref{eq:4-4}) and by ignoring
higher order terms in $\delta p_{i},$ it was shown in \cite{GA-VN_GA1980}
that the perturbations satisfy 
\begin{equation}
\frac{d}{dt}\delta p_{i}=\sum^{k}_{j=1}C_{ij}\delta p_{i},\label{eq:4-6}
\end{equation}
where $C_{ij}=\bar{p_{i}}\left(A_{ij}-\sum^{k}_{q=1}A_{qj}\bar{p_{j}}\right).$
Let $\mathbf{x}$ be an eigenvector of the matrix $\mathbf{C}=\left(C_{ij}\right)$,
and let $\lambda_{1},\lambda_{2},...,$$\lambda_{k}$ be the characteristics
roots of $\mathbf{C}$. The non-trivial solutions are obtained by
solving the characteristic equation $\det\left(\mathbf{Cx}-\lambda\mathbf{I}\right)=0$,
where $\mathbf{I}$ is the $k\times k$ corresponding identity matrix.
The steady-state of the model (\ref{eq:4-6}) is stable if all the
non-zero eigenvalues of $\mathbf{C}$ have negative real parts.

\subsection{Invasions and Invading Genotypes}

The payoff matrices that we studied above were found to be helpful
in understanding invasion of genotypes. Impact of the invasion of
a new genotype at steady-state is studied by considering a coefficient
$B_{ij},$ defined by 
\begin{equation}
B_{ij}=A_{ij}-A_{jj}.\label{eq:4-7}
\end{equation}
It was concluded that when $p_{i}\rightarrow0$, a population consisting
of $k$ genotypes is uninvadable by a new genotype, if, and only if,
$m_{i}<\bar{m}.$ The rate of change of mean fitness was obtained
from the equation: 
\begin{equation}
\frac{d\bar{m}}{dt}=\sum^{k}_{i=1}p_{i}(m_{i}-\bar{m})^{2}+\frac{1}{2}\sum^{k}_{i,j=1}(A_{ij}-A_{ji})p_{i}p_{j}(m_{i}-m_{j}).\label{eq:4-8}
\end{equation}
For the two-genotype case, stability conditions were found to be determined
by the signs of $B_{12}$ and $B_{21}.$ Under the two-genotype scenario,
the evolutionary stable composition is same as evolutionary stable
strategy. Such generalizations were not found to be valid for the
three-genotype cases.

Gadgil and others' research in this article paved the way for further
developments in evolutionary game dynamics, for instance refer to
\cite{Hines1987,Joshi1987,VSB1998,VN-SS2011,Gorebook}.

\section{\textbf{Harvesting, Refugia, and Dynamic Modeling}}

In a premarket economy, agricultural societies adopted harvesting
strategies without any formal awareness of ecological science. These
societies harvested based on simple rules, without any idea of the
long-term equilibrium effects on yields or weather forecasts. Gadgil,
along with Joshi in \cite{NVJ-MG1990} built a dynamic population
model, studied harvesting as a concave function of effort, and showed
that yields exhibit diminishing returns.

Let $B_{t}$ be the resource biomass, and let $r$ ($r>1)$ be the
intrinsic growth rate. First, a logistic model was used to express
biomass after growth and before harvesting, followed by other model
forms. Let this biomass be denoted by $B'_{t}$, represented by 
\begin{equation}
B'_{t}=rB_{t}(1-B_{t}).\label{eq:5-1}
\end{equation}

The harvest, $H_{t}$ depends on harvesting effort, $E_{t}$ and is
related to the available biomass after growth through an expression
\begin{equation}
H_{t}=B'_{t}(1-e^{-E_{t}}).\label{eq:5-2}
\end{equation}
This concave effort-harvesting relationship is assumed to have diminishing
marginal returns; see also \cite{Clark1985}. The resource biomass
at time $t+1$, say $B_{t+1}$, after the harvest is expressed as
\begin{equation}
B_{t+1}=B'_{t}e^{-E_{t}}.\label{eq:5-3}
\end{equation}

\subsection{Constant and Variable Harvesting}

Societies are assumed to move towards maximum sustainable yields (MSY).
Under the constant harvesting effort $E$, i.e. $E_{t}=E$, the resource
biomass $B^{*}$, and harvest equilibrium $H^{*}$ are given by 
\begin{equation}
B^{*}=1-\frac{e^{E}}{\ln r}\text{ for }E<\ln r,\label{eq:5-4}
\end{equation}
\begin{equation}
H^{*}=B^{*}(1-e^{-E}),\label{eq:5-5}
\end{equation}
achieve maximum MSY at $\hat{E}$, a unique effort level. For $E\geq\ln r$,
the extinction of the resource population occurs. The harvest attains
a maximum value of $\frac{(r-1)^{2}}{4r}$ at $\hat{E}=\ln\frac{1+r}{2}.$
Under the assumption that harvesters adjust their efforts from time
to time, the discrete version for change in the efforts, $D_{t}$
is represented by 
\begin{equation}
D_{t}=\left\{ \begin{array}{c}
xD_{t-1}\text{ if }H_{t}>H_{t-1}\\
-xD_{t-1}\text{ if }H_{t}<H_{t-1}
\end{array}\right.\label{eq:5-6}
\end{equation}
for a positive scaling factor $x$. The short-term harvest changes
can be discretely understood by $E_{t+1}=E_{t}+D_{t}.$

\subsection{Refugium Model}

Under the assumption that some fraction $a\in[0,1]$ of the resource
is always protected from harvesting, the basic model for $H$ is extended
here to represent 
\begin{equation}
H_{t}=(1-a)B'_{t}(1-e^{-E_{t}}),\label{eq:5-7}
\end{equation}
and the biomass remaining after exploitation of resources, $B_{t+1}$
is represented by 
\begin{equation}
B_{t+1}=(1-a)B'_{t}e^{-E_{t}}+aB'_{t}.\label{eq:5-8}
\end{equation}
The ideas of refugia provide constraints independent of harvester
behavior, and when $a>\frac{1}{r}$, the population cannot go extinct
regardless of the values of effort $E.$ When spatial mixing is effective,
refugia do not reduce optimal yields.

\section{\textbf{Forecasting Models for Spatially Structured Populations}}

Forecasting spatial structures of populations has value in understanding
resource exploitation and ecological conservation. Gadgil, along with
Xu, Boyce, Nanjundiah evaluated the performance of nonlinear forecasting
methods when populations exhibit chaotic and periodic behavior \cite{Xuetal2005}.
Multiple spatial configurations are examined to study dynamics of
each subpopulation within a population by assuming that subpopulations
follow Moran-Ricker nonlinear discrete-time model 
\begin{equation}
N_{t+1}=N_{t}\exp\left[r\left(1-\frac{N_{t}}{K}\right)\right],\label{eq:6-1}
\end{equation}
where $N_{t}$ is the density of a subpopulation at $t,$ $r$ is
the intrinsic growth rate, and $K$ is the carrying capacity. Further
information on such models in ecology can be found elsewhere, for
example, see in \cite{Bonsall,Moran,Ricker1954,ZhengKurse,Muni2025}.
Such framework was also applied to understand spatial structures of
prey-predator dynamics \cite{Molina,bapan}. Gadgil and others (Xu,
et. al. \cite{Xuetal2005}) contribution conclude that forecasting
in spatially structured systems are challenging and there are fundamental
limits to it. By considering a dispersal rate $D_{ij}$ from subpopulation
$i$ to subpopulation $j$ among $n$ subpopulations, a dispersal
model of the form: 
\begin{equation}
N'_{i,t}=N_{i,t}+\sum^{n}_{j=1\text{ }(j\neq i)}\left(D_{ji}N_{j,t}-D_{ij}N_{i,t}\right),\label{eq:6-2}
\end{equation}
where $D_{ij},D_{ji}<\frac{1}{n},$ and $N_{i,t}$ and $N'_{i,t}$
are the densities of subpopulation $i$ at time $t$ before and after
dispersal, respectively. For a density-dependent dispersal, they considered
a quadratic version of (\ref{eq:6-2}) expressed as 
\begin{equation}
N'_{i,t}=N_{i,t}+\sum^{n}_{j=1\text{ }(j\neq i)}\left(D_{ji}N^{2}_{j,t}-D_{ij}N^{2}_{i,t}\right).\label{eq:6-3}
\end{equation}
Using an artificial neural network framework, they examined environmental
noise. Let $\epsilon_{t}$ be the environmental noise in the system
assumed to be uniformly distributed within a specified interval. Incorporating
the noise, the model in (\ref{eq:6-1}) is written as 
\begin{equation}
N_{t+1}=N_{t}\exp\left[r\left(1-\frac{N_{t}}{K}\right)\right](1+\epsilon_{t}).\label{eq:6-4}
\end{equation}
Let $D=D_{ij}=D_{ji}.$ The dynamics at a critical value $D_{c}$
were studied. When $D<D_{c}$, aggregated populations lose deterministic
structure and subpopulations are asynchronous. When $D>D_{c}$, the
aggregated dynamics recover predictability, and subpopulations synchronize.

\section{\textbf{Conclusions and Discussion}}

We reviewed Madhav Gadgil's legacy in life-history theory. These studies
suggest a common approach to ecological theory: identify a biological
constraint, make the consequences of alternative strategies explicit
and examine how feedbacks change those consequences. Gadgil's contributions
connect age-specific allocation, interactions among genotypes, resource
exploitation and spatial prediction \cite{GA-Bos1970,GA-VN_GA1980,NVJ-MG1990,Xuetal2005}.
Gadgil's work emphasized that ecological outcomes depend on the conditions
under which a strategy is expressed. Reproductive effort changes subsequent
survival and growth, the composition of a population changes the success
of its constituent genotypes and harvesting changes the resource on
which future decisions depend.

The Gadgil-Bossert model made this reasoning explicit by connecting
reproductive effort to present fecundity and future demographic contributions
\cite{GA-Bos1970}. Its continuing value lies in the biological relationships
that must be specified before an optimal life history can be identified.
A useful direction is to estimate these relationships across environments
and test whether the same allocation model predicts both age-specific
reproduction and population growth. Such tests should distinguish
changes in resource acquisition from changes in allocation: individuals
with greater resources need not display the trade-offs expected when
resource availability is held constant \cite{VanNoordwijkDeJong1986,jaggi2024density}.
Jaggi et al. found changing population density alters life-history
tradeoffs in a food-limited population of Soay sheep \cite{jaggi2024density},
laying emphasis on the dynamic nature of tradeoffs and constraints.

It is also important to consider what traits are being optimized.
The population growth rate and lifetime reproductive output are distinct
metrics and their suitability as evolutionary criteria depends on
ecological regulation. Mylius and Diekmann \cite{MyliusDiekmann1995}
showed that different forms of density dependence can lead to different
optimization criteria. Gadgil and Solbrig's emphasis on resource allocation
provides a useful starting point for revisiting $r$- and $K$-selection
\cite{Ga-Solb}. A key question is how particular forms of crowding
alter survival, growth, and reproduction, and thereby change selection
on allocation. Treating $r$ and $K$ as universal endpoints would
obscure those mechanisms and the conditions under which different
traits can be optimized.

Gadgil, Nanjundiah and Gadgil \cite{GA-VN_GA1980} examined the dynamics
of interacting genotypes and defined evolutionarily stable compositions
using both resistance to changes in resident frequencies and resistance
to invasion. These are distinct tests: the response of existing types
to a perturbation does not by itself determine the outcome of introducing
a new type. Their framework therefore connects the maintenance of
diversity to the interactions that determine relative fitness. A natural
extension is to ask whether compositions stable under fixed conditions
retain these properties when resource availability or interaction
strengths change. This would connect equilibrium predictions to the
ecological conditions under which diversity persists.

Environmental variation offers another direction for extending the
allocation framework. Stochastic demographic theory shows that long-run
fitness can depend on temporal variability and autocorrelation as
well as average vital rates. One concrete question is whether allocation
strategies favoured under a constant resource supply remain favoured
when the same mean supply arrives in brief episodes, similar to the
pulse-press framework analyzed in \cite{jaggi2026transient}. Comparing
these environments could separate the effects of average conditions,
temporal predictability, and the speed at which allocation can adjust.
This extension would also require an explicit distinction between
frequency-dependent selection and plastic responses within an individual's
lifetime.

The harvesting work by Joshi and Gadgil \cite{NVJ-MG1990} linked
harvesting decisions to resource dynamics and examined refugia as
a means of maintaining exploited populations. Here, changes in behaviour
concern decisions by harvesters. The theoretical connection is the
feedback between present use and future opportunities. Their analysis
motivates a distinction between maximizing an equilibrium yield and
maintaining a resource under changing effort. Prediction must also
be matched to spatial scale. In the models examined by Xu and colleagues
\cite{Xuetal2005}, asynchronous subpopulations could produce aggregate
dynamics that were less amenable to nonlinear forecasting than local
dynamics. The relationship depended on dispersal and spatial configuration.
Gadgil \cite{gadgil1971dispersal} also examined analytically how
variation in habitat quality across space and time shapes the evolution
of dispersal. Further work analyzed when selection favors lower dispersal
rates in patchy landscapes \cite{kirkland2006dispersal} and when
temporal variability can favor migration between habitats \cite{wiener1994migration,jaggi2024temporal}.

These questions also provide a connection to Gadgil's participatory
biodiversity work. The People's Biodiversity Registers documented
local ecological knowledge and resource-use practices through collaboration
with communities and local institutions. The work has implications
for theoretical biology: knowledge of local resources, seasonal changes
and harvesting practices can help determine which mechanisms a model
should represent and which predictions can be tested. Continuing Gadgil's
approach thus calls for biologically and locally grounded assumptions,
considering forms of stability and density-dependence and a dialogue
between mathematical analysis and observation. Developing mathematical
theory that enhances our understanding of ecological and social conditions
will be an honor to Madhav Gadgil's legacy.

\clearpage{}

\end{document}